\documentclass{cas-dc}
\usepackage[utf8]{inputenc}
\usepackage[numbers]{natbib}
\usepackage{bm} 

\begin{document}

\let\WriteBookmarks\relax
\def\floatpagepagefraction{1}
\def\textpagefraction{.001}
\shorttitle{Height-averaged Navier-Stokes solver for hydrodynamic lubrication}
\shortauthors{H. Holey et~al.}

\title[mode = title]{Height-averaged Navier-Stokes solver for hydrodynamic lubrication}                      

\author[1,2]{Hannes Holey}  
\ead{hannes.holey@kit.edu}
\author[2,3]{Andrea Codrignani}
\author[1,3]{Peter Gumbsch}
\author[2,4]{Lars Pastewka}
\ead{lars.pastewka@imtek.uni-freiburg.de}

\address[1]{Institute for Applied Materials, Karlsruhe Institute of Technology, Straße am Forum 7, 76131 Karlsruhe, Germany}
\address[2]{Department of Microsystems Engineering (IMTEK), University of Freiburg, Georges-Köhler-Allee 103, 79110 Freiburg, Germany}
\address[3]{Fraunhofer IWM, Wöhlerstraße 11, 79108 Freiburg, Germany}
\address[4]{Cluster of Excellence livMatS, Freiburg Center for Interactive Materials and Bioinspired Technologies, University of Freiburg, Georges-K\"ohler-Allee 105, 79110 Freiburg, Germany}


\begin{abstract}
Modelling hydrodynamic lubrication is crucial in the design of engineering components
as well as for a fundamental understanding of friction mechanisms. The cornerstone of thin-film flow modelling is the Reynolds equation -- a lower-dimensional representation of the Stokes equation.
However, the derivation of the Reynolds equation is based on assumptions and fixed form constitutive relations, that may not generally be valid, especially when studying systems under extreme conditions.
Furthermore, these explicit assumptions about the constitutive behaviour of the fluid prohibit applications in a multiscale scenario based on measured or atomistically simulated data.
Here, we present a method that considers the full compressible Navier-Stokes equation in a height-averaged sense for arbitrary constitutive relations. We perform numerical tests by using a reformulation of the viscous stress tensor for laminar flow to validate the presented method comparing to results from conventional Reynolds solutions. The versatility of the method is shown by incorporating models for mass-conserving cavitation, wall slip and non-Newtonian fluids. 
This allows testing of new constitutive relations that not necessarily need to take a fixed form, and may be obtained from experimental or simulation data.
\end{abstract}

\begin{keywords}
Hydrodynamic Lubrication \sep Cavitation \sep Non-Newtonian Fluids \sep Finite-Volume Method
\end{keywords}

\ExplSyntaxOn
\keys_set:nn { stm / mktitle } { nologo }
\ExplSyntaxOff
\maketitle

\section{Introduction}

Although the fundamental equation for hydrodynamic lubrication was derived by \citet{reynolds1886_iv} almost 140 years ago, accurate descriptions of thin film fluid flow are still a research topic of ongoing interest. 
Major challenges in modelling arise in elastohydrodynamic lubrication (EHL), where a plethora of effects, such as non-Newtonian fluid behavior, surface roughness, wall slip or cavitation, has to be considered.
An extensive overview of the field of EHL is given in the review articles by \citet{lugt2011_review} and \citet{gropper2016_hydrodynamic}, the latter focussing on textured surfaces.

The derivation of the Reynolds equation is based on an asymptotic analysis of the incompressible Navier-Stokes equation under the assumption that the gap height is small compared to the lateral dimensions. As a result, inertial and body force terms can be neglected and the fluid pressure does not depend on the gap height coordinate. 
This allows integration of the momentum equations in the gap height coordinate to obtain the flow velocity profiles. Integration of the mass balance using these velocity profiles leads to the Reynolds equation for incompressible and isoviscous fluids \cite{hamrock2004_fundamentalsa}.

However, fluid compressibility and pressure-dependent viscosity cannot be ignored under severe loading conditions such as in EHL contacts. It is therefore common practice to introduce constitutive relations for the pressure-dependent density and viscosity to the asymptotic analysis \textit{a posteriori}. Yet, this can lead to wrong predictions for piezoviscous fluids, where the viscosity strongly depends on pressure \citep{rajagopal2003_inconsistency, bayada2013_new, gustafsson2015_nonlinear}.

Recently, \citet{almqvist2019_new} presented a compressible Reynolds equation for pressure-density relations that take the form of a power law. This includes the isothermal ideal gas law, the constant bulk modulus equations of state, as well as the Dowson-Higginson equation \citep{dowson1962_generalized} for mineral base oils. They found that except for the ideal gas, lower-dimensional formulations of the Navier-Stokes equations can be derived. For the ideal gas, the additional assumption of negligible inertial and body force terms has to be made, to integrate the momentum equations.

In this paper, we present a method that aims at overcoming the ambiguity in the choice of governing equation by using the full compressible Navier-Stokes equation in a height-averaged sense. The scheme is formulated for arbitrary constitutive relations, and is therefore applicable to complex lubrication problems.

We demonstrate the validity of the presented height-averaged Navier-Stokes equation for laminar flow problems. This means, we borrow the assumption of constant cross-film density from the derivation of the Reynolds equation and express the Newtonian viscous stress tensor as a function of conserved variables. The obtained pressure and density profiles of our transient finite volume solution are compared to various results from the literature using Reynolds-based solution schemes. Moreover, the formulation in terms of conserved variables allows a flexible implementation of cavitation models through the equation of state. 
Navier slip boundary conditions allow the investigation of lubrication with heterogeneous surface wettability, and non-Newtonian behavior is considered by commonly used semi-empirical models for shear-thinning and piezoviscous fluids.

\section{Governing equations}

In the following section, we propose a general method for thin film flows in lubrication, that considers the full compressible Navier-Stokes equation, but in a height-averaged sense. In order to validate the method, we will re-introduce assumptions that are typically found in the derivation of the Reynolds equation in section~\ref{sec:stress_model}.

\subsection{Height-averaged Navier-Stokes equation}

The local form of mass and momentum balance is given by the partial differential equation
\begin{equation}\label{eq:pde}
    \partial_t \bm{q} = - \mathrm{div}\left[\mathbb{f}(\bm{q})\right],
\end{equation}
where $\bm{q}$ denotes the 4-vector of densities of conserved variables (mass and momentum) and $\mathbb{f}(\bm{q})$ denotes the corresponding flux functions, represented by a $4\times 3$ matrix
\begin{align}\label{eq:def_q_flux}
\bm{q} = 
\begin{pmatrix}
\rho\\
\vec{j}
\end{pmatrix},
\quad
\mathbb{f}(\bm{q}) =
\begin{pmatrix} 
\vec{j} \\ 
\frac{1}{\rho}(\vec{j}\otimes\vec{j}) + p\underline{1} - \underline{\tau}(\bm{q}) 
\end{pmatrix}.
\end{align}
Note that the divergence operator in Eq.~\eqref{eq:pde} acts on the column index $[\mathrm{div}(\mathbb{A})]_i = \partial_j A_{ij}$. The first row of $\mathbb{f}$ contains the components of the mass flux vector $\vec{j}$, representing mass transport. Momentum transport is encoded in the convective acceleration term $(\vec{j}\otimes\vec{j})/\rho$, where $\otimes$ defines the outer product between two vectors as $[\vec{a}\otimes\vec{b}]_{ij} = a_i b_j$, and the pressure term $p\underline{1}$, where $\underline{1}$ denotes the $3\times 3$ unit matrix.

Pressure in a compressible fluid is in general a function of mass density and temperature. This equation of state is needed to close the system of balance laws.  Here, we limit ourselves to isothermal conditions, which is why we also do not include the energy balance equation, and therefore $p=p(\rho)$. Finally, the viscous stress tensor $\underline{\tau}$ is a function of the conserved variables, which is not further specified here. In section~\ref{sec:stress_model}, we show how to express the Newtonian viscous stress tensor as a function of $\bm{q}$ for laminar flow, which is later used for the numerical tests shown in section~\ref{sec:results}.

\begin{figure}
    \centering
    \includegraphics{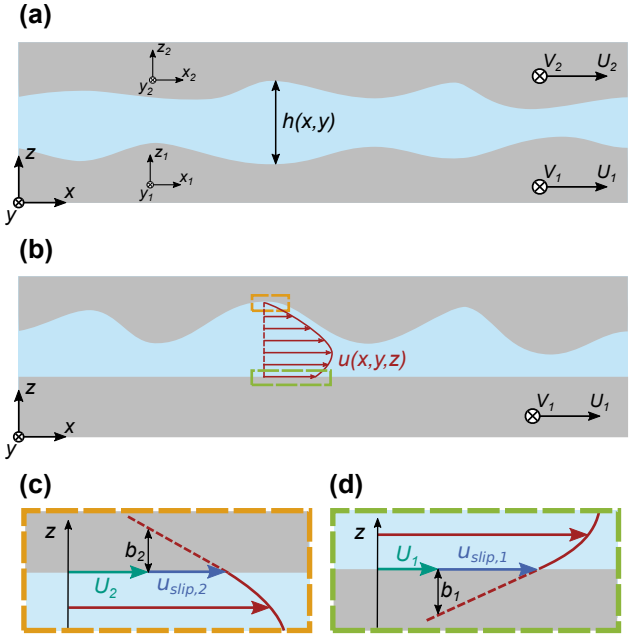}
    \caption{Sketch of the lubrication gap with height profiles of the lower and upper surface given in the local coordinate systems $(x_i,\,y_i,\,z_i)$, moving with constant velocities $\vec{U_i} = (U_i, V_i, 0)^\top$, $i\in[1,\,2]$ (a). The source term in Eq.~\eqref{eq:pde_avg} can be simplified by mapping the combined height profile on a stationary wall, while moving the flat counter surface (b). Deviation from no-slip boundary conditions is considered through the Navier slip length on the upper (c) and lower surface (d).
    }
    \label{fig:geometry}
\end{figure}

To arrive at a two-dimensional description we take an average in $z$-direction
\begin{align}\label{eq:pde_avg}
    \frac{1}{h} \int_{h_1}^{h_2} \partial_t \bm{q} \mathrm{d}z= - \frac{1}{h} \int_{h_1}^{h_2} (\partial_x\bm{f}_x + \partial_y\bm{f}_y + \partial_z\bm{f}_z) \mathrm{d}z,
\end{align}
where $\bm{f}_i$ denotes the $i$-th column of the flux function matrix $\mathbb{f}$, $h_1(x_1, y_1)$ and $h_2(x_2, y_2)$ are the height profiles of the lower and upper surface respectively, and their difference is denoted with $h$ as shown in Fig.~\ref{fig:geometry}a. Note that these profiles are expressed in coordinate systems that are attached to the walls, which move relative to the global reference frame with constant velocities $\vec{U}_1=(U_1, V_1, 0)^\top$ and $\vec{U}_2=(U_2, V_2, 0)^\top$.

Pulling the differentiation out of the integral yields for the l.\,h.\,s.
\begin{align}
\int_{h_1}^{h_2} \partial_t \bm{q} \mathrm{d}z &= 
\partial_t \int_{h_1}^{h_2}  \bm{q}\mathrm{d}z
- \bm{q}|_{z=h_2}\frac{\mathrm{d}h_2}{\mathrm{d}t}
+ \bm{q}|_{z=h_1}\frac{\mathrm{d}h_1}{\mathrm{d}t} \nonumber \\
&= 
h \partial_t \bar{\bm{q}} +
(\bar{\bm{q}} - \bm{q}|_{z=h_2})\frac{\mathrm{d}h_2}{\mathrm{d}t} 
- (\bar{\bm{q}} - \bm{q}|_{z=h_1})\frac{\mathrm{d}h_1}{\mathrm{d}t},
\end{align}
and for the first two terms on the r.\,h.\,s.
\begin{align}
\int_{h_1}^{h_2} &\partial_x \bm{f}_x \mathrm{d}z =
\partial_x \int_{h_1}^{h_2}  \bm{f}_x \mathrm{d}z
- \bm{f}_x|_{z=h_2}\frac{\partial h_2}{\partial x}
+ \bm{f}_x|_{z=h_1}\frac{\partial h_1}{\partial x} \nonumber \\
&=
h \partial_x \bar{\bm{f}}_x +
(\bar{\bm{f}}_x - \bm{f}_x|_{z=h_2})\frac{\partial h_2}{\partial x} 
- (\bar{\bm{f}}_x - \bm{f}_x|_{z=h_1}) \frac{\partial h_1}{\partial x}, \\
\int_{h_1}^{h_2} &\partial_y \bm{f}_y \mathrm{d}z = 
\partial_y \int_{h_1}^{h_2}  \bm{f}_y \mathrm{d}z
- \bm{f}_y|_{z=h_2}\frac{\partial h_2}{\partial y}
+ \bm{f}_y|_{z=h_1}\frac{\partial h_1}{\partial y} \nonumber \\
&=
h \partial_y \bar{\bm{f}}_y +
(\bar{\bm{f}}_y - \bm{f}_y|_{z=h_2})\frac{\partial h_2}{\partial y} 
- (\bar{\bm{f}}_y - \bm{f}_y|_{z=h_1}) \frac{\partial h_1}{\partial y},
\end{align}
where overbars denote height-averages $\bar{\phi} = \frac{1}{h}\int_{h_1}^{h_1}\phi\,\mathrm{d}z$.
The third term under the integral on the r.\,h.\,s. of Eq.~\eqref{eq:pde_avg} can be evaluated directly at the fluid-wall boundaries
\begin{align}
\frac{1}{h}\int_{h_1}^{h_2} \partial_z \bm{f}_z \mathrm{d}z = 
\frac{1}{h} \left( \bm{f}_z|_{z=h_2} - \bm{f}_z|_{z=h_1}\right).
\end{align}
Hence, the averaged scheme is composed of a two-dimensional divergence operator acting on height-averaged flux functions and of terms containing both averaged and unaveraged flux functions, the latter evaluated at the top and bottom walls. The terms outside of the divergence operator can be regarded as geometrical source terms due to the reduction of dimensionality.

The averaged scheme reads
\begin{align}\label{eq:pde_avg_final}
  \partial_t \bar{\bm{q}} = - \partial_x \bar{\bm{f}}_x - \partial_y\bar{\bm{f}}_y - \bm{s},
\end{align}
with
\begin{align}
\bm{s} 
&= \frac{1}{h}\Bigl(
 \frac{\partial h_2}{\partial x}(\bar{\bm{f}}_x - \bm{f}_x|_{z=h_2})
-\frac{\partial h_1}{\partial x}(\bar{\bm{f}}_x - \bm{f}_x|_{z=h_1}) \nonumber \\
&\mathrel{\phantom{=\frac{1}{h}}}
+\frac{\partial h_2}{\partial y}(\bar{\bm{f}}_y - \bm{f}_y|_{z=h_2})
-\frac{\partial h_1}{\partial y}(\bar{\bm{f}}_y - \bm{f}_y|_{z=h_1}) \nonumber \\
&\mathrel{\phantom{=\frac{1}{h}}}
-\frac{\mathrm{d} h_2}{\mathrm{d} t}(\bar{\bm{q}} - \bm{q}|_{z=h_2})
+\frac{\mathrm{d} h_1}{\mathrm{d} t}(\bar{\bm{q}} - \bm{q}|_{z=h_1}) \nonumber \\
&\mathrel{\phantom{=\frac{1}{h}}}
+ \bm{f}_z|_{z=h_2} - \bm{f}_z|_{z=h_1}\Bigr).
\end{align}
The total time derivative for the upper ($i=2$) and lower ($i=1$) rigid surfaces can be written as
\begin{equation}
    \frac{\mathrm{d} h_i}{\mathrm{d} t} = \frac{\partial h_i}{\partial x}U_i + \frac{\partial h_i}{\partial x}V_i.
\end{equation}
Without loss of generality, we assume that the lower wall is flat ($h_1 = \mathrm{const.}$) and the upper wall is stationary ($\vec{U}_2 = \vec{0}$) as shown in Fig.~\ref{fig:geometry}b. Then, the source term simplifies to
\begin{align}
   \bm{s} 
&= \frac{1}{h}\Bigl(
 \frac{\partial h}{\partial x}(\bar{\bm{f}}_x - \bm{f}_x|_{z=h_2})
+\frac{\partial h}{\partial y}(\bar{\bm{f}}_y - \bm{f}_y|_{z=h_2}) \nonumber \\
&\mathrel{\phantom{=\frac{1}{h}}}
+ \bm{f}_z|_{z=h_2} - \bm{f}_z|_{z=h_1}\Bigr). 
\end{align}

For flat channels, where the gradient $\partial_i h$ disappears, the source term only consists of momentum flux contributions in $z$-direction evaluated at the upper and lower wall, respectively. Since the walls are impenetrable for the liquid, mass flux in $z$-direction is zero there. Moreover, if the walls are at rest, i.\,e. for pure Poiseuille flow, the source term vanishes completely.

The presented scheme is entirely formulated in terms of the densities of conserved variables. 
Given an equation of state $p(\rho)$, the hyperbolic flux components are automatically defined in terms of $\bm{q}$ by definition. However, the viscous stress tensor is typically given as a function of the velocity field $\vec{u}$. In the next section, we describe how the viscous stress tensor for laminar flow can be cast into a functional form that takes conserved variables as arguments.

\subsection{Viscous stress tensor in conserved variables}
\label{sec:stress_model}

In fluid dynamics calculations, the viscous stress tensor is usually expressed as a function of the velocity field $\vec{u}$ and the system is discretized in terms of the primitive variables $\vec{u}$, $p$ and $T$. 
Here, as described above, we need to express the viscous stress tensor as a function of height-averaged conserved variables $\bar{\bm{q}}$.
The viscous stress tensor for a compressible Newtonian fluid is given by
\begin{align}\label{eq:stress_newton}
    \underline{\tau} = \eta \left(\nabla \vec{u} + (\nabla \vec{u})^\top \right)+ (\zeta -\frac{2}{3}\eta) \, (\nabla \cdot \vec{u})\,\underline{1},
\end{align}
where $\eta$ and $\zeta$ are the coefficients of shear and bulk viscosity respectively. In order to express the components of the viscous stress tensor $\underline{\tau}$ in terms of $\bar{\bm{q}}$, we choose a second degree polynomial ansatz for the velocity field $\vec{u} = (u, v, w)^\top$, which shall be parameterized to represent Poiseuille- and Couette-type flow in streamwise direction 
\begin{subequations}
\begin{align}
u(z) &= \alpha_1 z^2 + \beta_1 z + \gamma_1, \\
v(z) &= \alpha_2 z^2 + \beta_2 z + \gamma_2, \\
w(z) &= 0.
\end{align}
\end{subequations}
The cross film velocity component $w(z)$ is assumed to be zero in this example of laminar flow, such that the momentum balance in $z$-direction does not have to be considered. However, a non-zero cross film velocity ansatz is generally possible.

The parameters $\beta_i$ and $\gamma_i,\;i=1,\,2$ are determined from the boundary conditions at the bottom and top wall, namely
\begin{subequations}
\begin{align}
u(h_1) &= U_1 + b_1(x,y)\left(\frac{\partial u}{\partial z}\right)_{z=h_1}, \\
u(h_2) &= U_2 - b_2(x,y)\left(\frac{\partial u}{\partial z}\right)_{z=h_2}, \\
v(h_1) &= V_1 + b_1(x,y)\left(\frac{\partial v}{\partial z}\right)_{z=h_1}, \\
v(h_2) &= V_2 - b_2(x,y)\left(\frac{\partial v}{\partial z}\right)_{z=h_2},
\end{align}
\end{subequations}
where $b_1(x,y)$ and $b_2(x,y)$ are the Navier slip length \citep{navier1823memoire} of the upper and lower surface, respectively.

Surface slip becomes relevant for confined lubricants, when the gap height is in the order of the slip length, such as in EHL contacts or in the boundary lubrication regime. The concept of slip length arises from the assumption that the interfacial shear stress is proportional to the slip velocity with the constant of proportionality quantifying the friction between fluid and solid. Using the Newtonian fluid constitutive equation, the slip length can therefore be described as a combination of a purely bulk property (shear viscosity) with a purely interfacial one (friction) \citep{bocquet2007_flow}. Geometrically, the slip length can be interpreted as the subsurface distance where the fluid velocity would equal that of the wall when linearly extrapolated, as shown in Fig.~\ref{fig:geometry}c and \ref{fig:geometry}d.

Naturally, many surfaces have heterogeneous surface chemistry, or surface properties can be tailored into particular stick-slip patterns, and therefore the slip length is a function of the lateral coordinates.
In the remainder of this paper, we assume that slip is only present at the top wall ($b_1(x,y) = 0$) and write $b(x,y)=b_2(x,y)$ for simplicity.

Furthermore, we assume constant mass density across the channel height, which is typically found after the asymptotic analysis in the derivation of the Reynolds equation, allowing to integrate the momentum equations. We want to emphasize, that these rather heuristic arguments are only introduced to be able to validate our method against standard Reynolds solutions, but do not limit its generality. Other expressions for the velocity and density profiles may also be possible.

Using the aforementioned assumption and the definition of the height-averaged mass flux
\begin{align}\label{eq:avg_flux}
\bar{\vec{j}} &= \frac{\rho}{h}\int_{h_1}^{h_2}\vec{u}\,\mathrm{d}z,
\end{align}
we obtain the remaining parameters $\alpha_i$, which describe the Poiseuille contribution to the flow.
This enables us to describe the viscous stress tensor according to Eq.~\eqref{eq:stress_newton} as a function of the average mass flux and density.

Hence, all entries of the flux matrix are given as a function of averaged conserved variables $\mathbb{f} = \mathbb{f}(\bar{\bm{q}})$. Computing the source term also requires unaveraged flux values at the top and bottom wall. These are automatically defined by the choice of velocity and density profiles.
The components of the averaged viscous stress tensor (Eq.~\eqref{eq:tau_avg}) and the local components of the viscous stress tensor at the bottom and top wall (Eqs.~\eqref{eq:tau_bottom} and \eqref{eq:tau_top}) are given in the appendix.

\subsection{Numerics}

We use a finite volume discretization of the two-dimensional domain with an explicit time integration scheme to solve Eq.~\eqref{eq:pde_avg_final}. MacCormack's \cite{maccormack2003_effect} method is easy to implement and including source terms into the predictor-corrector scheme is straightforward. The discretized version of the system reads
\begin{subequations}
\begin{align}
\bm{Q}_{i,j}^\ast &= \bm{Q}_{i,j}^{n} -\frac{\Delta t}{\Delta x}(\bm{F}_{\bm{x};\,i+1,j}^n - \bm{F}_{\bm{x};\,i,j}^n)\nonumber \\
&\mathrel{\phantom{=}} - \frac{\Delta t}{\Delta y}(\bm{F}_{\bm{y};\,i,j+1}^n - \bm{F}_{\bm{y};\,i,j}^n) - \Delta t\bm{S}_{i,j}^n, \\
\bm{Q}_{i,j}^{\ast\ast} &= \bm{Q}_{i,j}^\ast -\frac{\Delta t}{\Delta x}(\bm{F}_{\bm{x};\,i,j}^\ast - \bm{F}_{\bm{x};\,i-1,j}^\ast)\nonumber \\
&\mathrel{\phantom{=}} - \frac{\Delta t}{\Delta y}(\bm{F}_{\bm{y};\,i,j}^\ast - \bm{F}_{\bm{y};\,i,j-1}^\ast) - \Delta t\bm{S}_{i,j}^\ast, \\
\bm{Q}_{i,j}^{n+1} &= \frac{1}{2}\left(\bm{Q}_{i,j}^{\ast\ast} + \bm{Q}_{i,j}^{n}\right),
\end{align}
\end{subequations}
where $\bm{Q}_{i,j}^\ast$ and $\bm{Q}_{i,j}^{\ast\ast}$ are intermediate solutions obtained from forward and backward differencing of the discretized flux vectors $\bm{F}_{(\bullet);\,i,j}^n$, and $\bm{S}_{i,j}^{(\bullet)}$ is the discrete source term. Here, subscripts $i$ and $j$ denote the grid cell with side lengths $\Delta x$ and $\Delta y$, the superscript $n$ denotes discrete time points separated by the time step $\Delta t$.
The order of differencing directions is arbitrary and one could also switch between backwards and forwards differences after every consecutive time step. Although the intermediate steps are first order accurate, the overall scheme is second order accurate in space and in time. 

Other Jacobian-free integration schemes such as Richtmyer's two-step version of the Lax-Wendroff scheme could also be implemented straightforwardly (see Ref.~\citep{zhang1999_modifications} for an overview of modifications to the Lax-Wendroff scheme with source terms). In the linear case (and without source term), MacCormack's method is equivalent to the Lax-Wendroff scheme \cite{leveque2002_finite}. 

Boundary conditions are implemented using a ghost cell approach, leading to $N+2$ cells in each direction, due to the three-point stencil of the MacCormack scheme. For Dirichlet type boundary conditions, the value of the ghost cell is chosen such that the prescribed value of the conserved variable is satisfied at the cell boundary by linear interpolation between the boundary cell and the ghost cell. Inflow and outflow boundary conditions $\nabla \bm{q} = \bm{0}$ are satisfied by setting the ghost cell value to the boundary cell value. Implementation of periodic boundary conditions is also straightforward. One-dimensional examples in section~\ref{sec:results} are computed using the full two-dimensional description with periodic boundary conditions in $y$-direction, leading to a $(N + 2) \times 3$-grid.

The scheme is implemented in Python using the \texttt{numpy} library. Compared to conventional stationary Reynolds solvers, computation time can be large, since we are resolving the full transient behavior of the flow until steady state is reached, while the time step is limited by the Courant-Friedrichs-Lewy condition. Computational speed-up is achieved by spatial domain decomposition in combination with a parallel implementation using the \texttt{mpi4py} bindings to the message passing interface (MPI).
The source code is made publicly available under the terms of the MIT license\footnote{\url{https://github.com/hannes-holey/hans}}.

\section{Lubricant models}
\label{sec:lub_models}

The scheme described above is tested for laminar thin film flow as described in section~\ref{sec:stress_model}. In the following section we describe the models we used for compressibility, cavitation and non-Newtonian viscosity, that lead to a more realistic description of hydrodynamic lubrication.
\subsection{Equation of state and cavitation}
The system of continuity equation and momentum balance needs to be closed by an equation of state, relating the pressure in the hyperbolic flux contribution to the mass density. Various types of equations describing the compressibility of the fluid can be used, ranging from the ideal gas to semi-empirical descriptions, such as the isothermal Dowson-Higginson \citep{dowson1962elasto, dowson1966_elastohydrodynamic} equation of state, 
\begin{align}\label{eq:eos_DH}
    p(\rho) = p_0 + C_1 \frac{\rho - \rho_0}{C_2 \rho_0 - \rho},
\end{align}
with fitting parameters $C_1$ and $C_2$, which is commonly used for mineral base oils.

However, such expressions do not reflect the effect of vaporous cavitation, which mainly occurs in diverging contact geometries, where the pressure may fall below the vapor pressure of the liquid. The first models that accounted for mass-conserving cavitation were based on the work of Jakobsson, Floberg, and Olsson (JFO) \cite{jakobsson1957_finite, floberg1960two, floberg1961lubrication, olsson1965cavitation}, who formulated boundary conditions for the rupture and reformation of a fluid film under the assumption, that the pressure is constant in the cavitated regions. Elrod's and Adam's cavitation algorithm \citep{elrod1975computer, elrod1981cavitation} builds upon the JFO approach, but added a switch function to the Reynolds equation that suppresses the Poiseuille contribution to the flow in the cavitated regions. The advantage of the Elrod-Adams (EA) approach is that a single equation can be used for the whole domain.

Here, we incorporate mass-conserving cavitation directly through the equation of state, either by fixing the pressure to a constant value $p_\mathrm{cav}$ at densities lower than the saturation density, which is conceptually similar to the EA algorithm, or by using a unique equation of state describing the behavior of vapor, liquid and vapor-liquid-mixture as in the model of  \citet{bayada2013_compressible}. 

\begin{figure}
    \centering
    \includegraphics{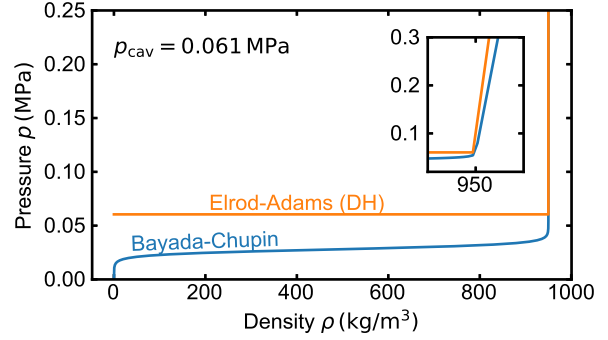}
    \caption{Equations of state (EOS) used for modelling mass-conserving cavitation in combination with the height-averaged Navier-Stokes solver. \citet{bayada2013_compressible} use a smooth function to interpolate between the pure vapor and liquid phases, whereas the Elrod-Adams algorithm with Dowson-Higginson EOS (cf. Ref~\citep{sahlin2007_cavitation}) leads to a sharp edge at the cavitation pressure.}
    \label{fig:eos}
\end{figure}

Their model assumes constant compressibility both in the liquid and the vapor phase, defined by the velocities of sound $c_\mathrm{l}$ and $c_\mathrm{v}$, respectively. The assumption that both phases of the vapor-liquid mixture have the same velocity allows to homogenize the dynamics of the flow in the cavitated region such that the same set of equations as in the full film region can be used. The pressure density relation for the mixture of vapor bubbles and liquid is defined in terms of the vapor fraction  $\alpha=(\rho-\rho_\mathrm{l})/(\rho_\mathrm{v}-\rho_\mathrm{l})$,  in which  $\rho_\mathrm{v}$ and $\rho_\mathrm{l}$ are the density of the vapor and the liquid at the wet point and the bubble point respectively. Bayada and Chupin used a relation proposed by \citet{wijngaarden1972one}
\begin{align}
    \frac{1}{c_\mathrm{f}^2} = \rho \left(\frac{\alpha}{c_\mathrm{v}^2\rho_\mathrm{v}} + \frac{1-\alpha}{c_\mathrm{l}^2\rho_\mathrm{l}}\right),
\end{align}
relating the fluid's velocity of sound $c_\mathrm{f}$ with the vapor fraction. Integrating the equation {$\mathrm{d}p/\mathrm{d}\rho=c_\mathrm{f}^2$} in each of the regions and requiring continuity of the pressure at the transition from vapor to mixture ($\alpha=1$) as well as from mixture to vapor ($\alpha=0$) and that $p(0) = 0$, one finds
\begin{align}\label{eq:bayada_eos}
p(\rho) = 
\begin{cases}
c_\mathrm{v}^2 \rho, &\alpha \geq  1, \\
p_\mathrm{cav} + (\rho - \rho_\mathrm{l})c_\mathrm{l}^2, &\alpha \leq 0, \\
p_\mathrm{vm} + N \ln\left( \frac{\rho_\mathrm{v} c_\mathrm{v}^2 \rho}{\rho_\mathrm{l}(\rho_\mathrm{v} c_\mathrm{v}^2 (1-\alpha)+\rho_\mathrm{l} c_\mathrm{l}^2 \alpha)}\right), &0< \alpha < 1,
\end{cases}
\end{align}
with pressure at the transition points from vapor to mixture
\begin{align}
p_\mathrm{vm} &= \rho_\mathrm{v} c_\mathrm{v}^2,
\end{align}
and from mixture to liquid (i.\,e. the cavitation pressure)
\begin{align}
p_\mathrm{cav} = \rho_\mathrm{v} c_\mathrm{v}^2 - N\ln\left(\frac{\rho_\mathrm{v}^2 c_\mathrm{v}^2}{\rho_\mathrm{l}^2 c_\mathrm{l}^2}\right), \quad
N = \frac{\rho_\mathrm{v}c_\mathrm{v}^2\rho_\mathrm{l}c_\mathrm{l}^2(\rho_\mathrm{v} -\rho_\mathrm{l})}{\rho_\mathrm{v}^2c_\mathrm{v}^2 - \rho_\mathrm{l}^2c_\mathrm{l}^2}.
\end{align}
Note that, for convenience, in Eq.~\eqref{eq:bayada_eos} the definition of $\alpha$ is extended to values larger than one, or smaller than zero. Obviously, the interpretation as the vapor fraction is then no longer valid in these regimes. Fig.~\ref{fig:eos} shows an example of Eq.~\eqref{eq:bayada_eos} for material parameters that lead to a cavitation pressure of $0.061\,\mathrm{MPa}$. For comparison, an EOS following the Dowson-Hingginson relation in the liquid phase and having constant pressure everywhere else is given, which represents a realization of the EA algorithm.

The pressure-density relation is accompanied by a viscosity model that interpolates between the vapor and liquid viscosity, the simplest one being a linear interpolation in terms of the vapor fraction $\alpha$
\begin{align}
    \eta(\alpha) = \eta_\mathrm{v} \alpha + (1 - \alpha) \eta_\mathrm{l}, \quad 0 \leq \alpha \leq 1.
\end{align}

\subsection{Non-Newtonian fluids}
Lubricants in the (elasto-)hydrodynamic lubrication regime are subject to high pressures and shear rates which alter the rheology of the fluid. To account for the rise in the lubricant's viscosity due to pressure increase and the shear-thinning behavior, various phenomenological models can be included, i.\,e. non-Newtonian flow is represented by a generalized Newtonian fluid approach. The pressure dependence of the viscosity is commonly modelled using an exponential form such as the Barus equation
\begin{align}\label{eq:barus}
    \eta(p) = \eta_0 \exp (\alpha p),
\end{align}
where $\alpha$ is the pressure viscosity coefficient, typically ranging between $0.1$ and $0.2\,\mathrm{GPa}^{-1}$. Although Eq.~\eqref{eq:barus} is known to overestimate the viscosity for higher pressures and more accurate empirical modifications exist, we will use it in this study for the sake of simplicity.

The two most prominent models for shear-thinning are the Eyring \cite{eyring1936_viscosity, ewell1937_theory} and Carreau \cite{carreau1972_rheological} models. The Eyring model assumes that molecular rearrangements due to external shear are thermally activated processes, in which the potential energy barriers to be overcome have equal heights. This leads to a shear-thinning behavior at large stresses and strain rates given by
\begin{align}\label{eq:eyring}
    \eta(\dot{\gamma}) = \frac{\tau_0}{\dot{\gamma}}\sinh^{-1}\left(\frac{\eta_0 \dot{\gamma}}{\tau_0}\right), 
\end{align}
with $\tau_0 = \eta_0 \gamma_0$ being a reference shear stress in the Newtonian regime and  shear rate $\dot{\gamma} = (\dot{\gamma}_{xz}^2 + \dot{\gamma}_{yz}^2)^{1/2}$. On the other hand, the Carreau model is based on the idea that there is not only a single, but rather a broad distribution of barrier heights. This leads to a power-law scaling of shear stress with rate, and hence, the Carreau equation reads
\begin{align}
    \eta(\dot{\gamma}) = \eta_0 \left(1 + (\lambda\dot{\gamma})^2\right)^\frac{n - 1}{2},
\end{align}
where $\lambda = 1/\dot{\gamma}_0$ is a characteristic time above which no shear-thinning happens and the scaling exponent $n$ is between $0$ and $1$. In a recent molecular dynamics study on model fluids under EHL conditions, \citet{jadhao2019_rheological} showed that there is a generic transition from power-law behavior at low pressure, and therefore low viscosity, to Eyring behavior at high pressure, where the viscosity is high and thermal energy becomes small compared to the shear barrier height.

Next to the piezoviscous and shear-thinning effects, temperature strongly affects the lubricant's viscosity. However, since we only investigate isothermal conditions, thermoviscous effects are left out of consideration.

\section{Numerical tests}\label{sec:results}

In this section we test the validity and versatility of the proposed height-averaged Navier-Stokes solver by adapting the lubricant models presented in section~\ref{sec:lub_models} and comparing the converged steady-state pressure and density profiles to literature results. 
All tests except the ones shown in section~\ref{sec:slip} employ no-slip boundary conditions, i.\,e. $b(x,y)=0$. Newtonian fluid behavior is assumed in all but the last example (section~\ref{sec:res_non-newton}), where we apply a generalized Newtonian fluid constitutive equation with shear rate and pressure-dependent viscosity.

\subsection{Inclined slider}

\begin{figure}
    \centering
    \includegraphics{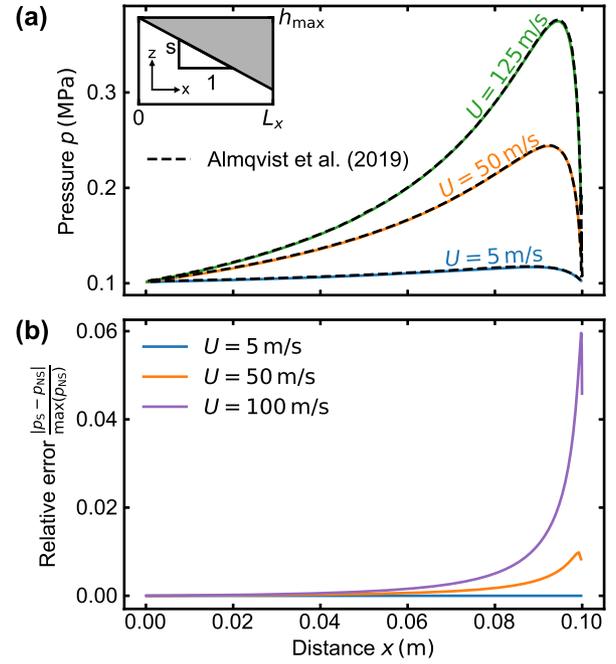}
    \caption{Pressure profiles for a gas-lubricated inclined slider geometry for three different sliding speeds and comparison to results from Ref.~\citep{almqvist2019_new} (a). The influence of including convective acceleration terms is studied in (b) for increasing sliding speed. In both cases, the fluid is described by an isothermal ideal gas equation of state and has constant viscosity.}
    \label{fig:slider1D_pressure_almqvist}
\end{figure}

As a first test for the transient numerical scheme, we use an inclined slider geometry with an ideal gas EOS and compare the steady state solution with the results of Ref.~\cite{almqvist2019_new} obtained from a compressible Reynolds equation. The height profile is given by $h(x) = h_\mathrm{max} - s x$ for $x\in[0,\,L]$, where $s$ is the slope of the pad. Here, we test the scheme at different sliding speeds $U$ for a bearing with $L = 0.1\,\mathrm{m}$, $h_\mathrm{max}=66\,\mathrm{\mu m}$ and $s=5.6\cdot10^{-4}$. The equation of state is given as 
\begin{align}
p(\rho) = \frac{p_0}{\rho_0}\rho,   
\end{align}
with ambient pressure $p_0 = 101325\,\mathrm{Pa}$ and ambient density $\rho_0 =1.1853\,\mathrm{kg/m^3}$. The shear viscosity is assumed to be constant at value of $\eta = 18.46\cdot10^{-6}\,\mathrm{Pa\,s}$, and the bulk viscosity $\zeta$ is set to zero. The obtained pressure profiles for three different velocities of the gas-lubricated slider bearing are shown in Fig.~\ref{fig:slider1D_pressure_almqvist}a. A substantial increase of load bearing capacity with sliding speed can be observed and the results match perfectly with the ones obtained in Ref.~\cite{almqvist2019_new}.
The simulation results for the inclined slider geometry, as well as all other tests in the remainder of this section, are obtained under the assumption that convective (nonlinear) inertial terms in the momentum equations can be neglected, which is a reasonable assumption in most cases. 
However, as shown in Ref.\citep{almqvist2019_new} for slider bearings and the ideal gas EOS, the Reynolds solution, although rigorously derived for power-law pressure-density relations, deviates from the full Navier-Stokes result with increasing sliding velocity.
Since the convective acceleration term is generally included into the definition of the flux function in Eq.~\eqref{eq:def_q_flux}, we have tested its influence for the inclined slider geometry for three different velocities. In Fig.~\ref{fig:slider1D_pressure_almqvist}b, the relative difference between the height-averaged Stokes solution (without nonlinear term, $p_\mathrm{S}$) and the height-averaged Navier-Stokes solution ($p_\mathrm{NS}$) is shown. The maximum pressure error due to neglection of inertial terms for a sliding speed of $U=100\,\mathrm{m/s}$ is 6\,\%. 

\begin{figure}
    \centering
    \includegraphics{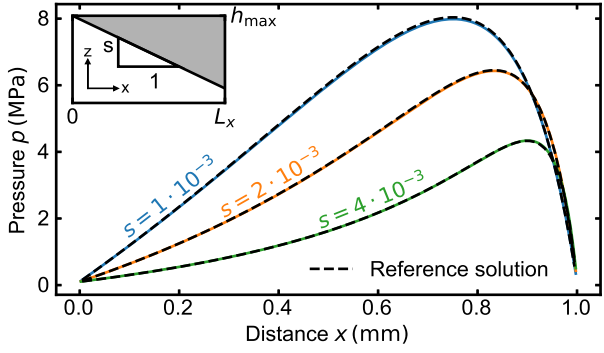}
    \caption{Pressure profiles for an inclined slider geometry with three different slopes at sliding speed $U = 0.1\,\mathrm{m/s}$ and Dowson-Higginson equation of state. Reference solutions are obtained from a finite difference solver for the compressible Reynolds equation used in Ref.~\citep{codrignani2017_scaling}.}
    \label{fig:slider1D_pressure}
\end{figure}

We further test the scheme for fluids with varying compressibility using the Dowson-Higginson EOS (Eq.~\eqref{eq:eos_DH}). Here, we choose the parameters reported in Ref.~\citep{sahlin2007_cavitation} ($C_1 = 2.22\,\mathrm{GPa}$, $C_2 = 1.66$) and a slider geometry with inlet gap height $h_\mathrm{max} = 1.5\,\mathrm{\mu m}$, sliding speed $U = 0.1\,\mathrm{m/s}$ and three different slopes. The obtained pressure profiles are shown in Fig.~\ref{fig:slider1D_pressure} and are compared with results obtained from the compressible Reynolds solver reported in \citet{codrignani2017_scaling}. As in the previous case, the obtained pressure profiles match the Reynolds reference solution.

\subsection{(Twin-) parabolic slider}
\begin{figure}
    \centering
    \includegraphics{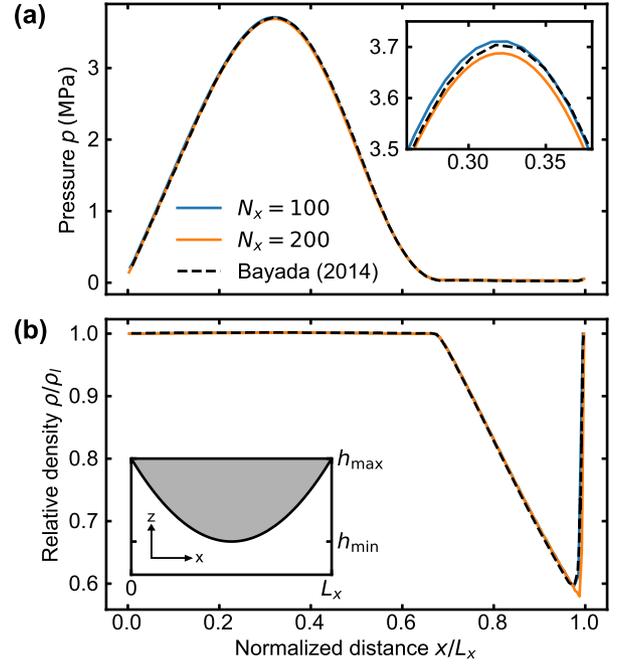}
    \caption{Pressure profile (a) and saturation profile (b) for the parabolic slider geometry and comparison to the results of Ref.~\cite{bayada2014_compressible}.}
    \label{fig:pslider1D_bayada}
\end{figure}

To study the effect of cavitation in hydrodynamic lubrication, we use a converging-diverging height profile, where we expect cavitation bubble formation in the diverging part of the bearing. Therefore, we use a parabolic slider geometry, which has been widely used to benchmark cavitation models. The height profile of the one-dimensional bearing is given by
\begin{align}
    h(x) = \frac{4(h_\mathrm{max} - h_\mathrm{min})}{L_x^2} \left(x-\frac{L_x}{2}\right)^2 + h_\mathrm{min},
\end{align}
with maximum gap height $h_\mathrm{max} = 50.8\,\mathrm{\mu m}$, minimum gap height $h_\mathrm{min} = 25.4\,\mathrm{\mu m}$, and length $L_x = 76.2\,\mathrm{mm}$. The parameters describing the lubricant through Eq.~\eqref{eq:bayada_eos} are given in Tab.~\ref{tab:bayada_params}. The resulting pressure profiles for two different discretizations ($N=100$ and $N=200$) are shown in Fig.~\ref{fig:pslider1D_bayada}a and the relative density or saturation $\rho/\rho_l$ is shown in Fig.~\ref{fig:pslider1D_bayada}b. Both agree very well with the profiles presented in Ref.~\cite{bayada2014_compressible}.

\begin{table}
    \caption{Parameters for the equation of state and the viscosity of liquid and vapor phase following Ref.~\citep{bayada2013_compressible}.}
    \centering
    \begin{tabular}{cccccc}
    \toprule
    $\eta_\mathrm{l}$ & $\eta_\mathrm{v}$ & $c_\mathrm{l}$ & $c_\mathrm{v}$  & $\rho_\mathrm{l}$ & $\rho_\mathrm{v}$\\
    $(\mathrm{Pa\,s})$ & $(\mathrm{Pa\,s})$ & $(\mathrm{m/s})$ & $(\mathrm{m/s})$  & $(\mathrm{kg/m^3})$ & $(\mathrm{kg/m^3})$\\
    \midrule
    $0.039$ & $3.9\cdot10^{-5}$ & $1600$ & $352$ & $850$ & $0.019$ \\
    \bottomrule
    \end{tabular}
    \label{tab:bayada_params}
\end{table}

The cavitation pressure $p_\mathrm{cav}$, or, in the notation of Bayada and Chupin, $p_\mathrm{ml}$, for the presented results is approximately $0.061\,\mathrm{MPa}$. As shown in their paper, the obtained results can be directly compared to solutions using the JFO/EA formalism, when using the same cavitation pressure and a constant compressibility EOS in the full film region.

\begin{figure}
    \centering
    \includegraphics{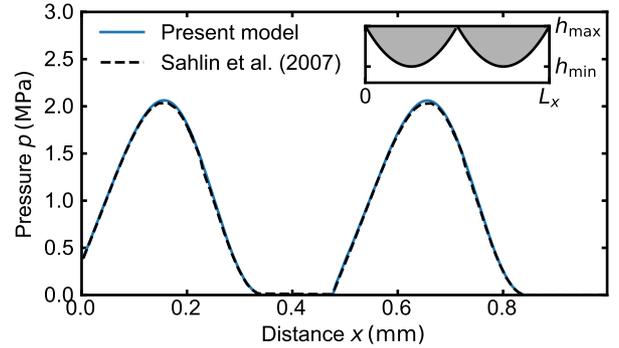}
    \caption{Pressure profile for the twin parabolic slider geometry and comparison to the results of Ref.~\cite{sahlin2007_cavitation}.}
    \label{fig:twin_pslider1D_sahlin}
\end{figure}

Here, we proceed with another numerical test using an EOS with varying compressibility in combination with the EA algorithm, as for instance presented in Ref.~\cite{sahlin2007_cavitation}. The so called twin-parabolic slider is used, consisting of two neighboring "bumps" with a parabolic shape and the same minimum and maximum gap heights as in the previous example. The length and sliding speed of the bearing are also the same as before. The pressure boundary conditions at the inlet and outlet are $p_\mathrm{in} = 3.36\,p_0$ and $p_\mathrm{out} = p_\mathrm{cav}$, respectively, with ambient pressure $p_0 = 10^5\,\mathrm{Pa}$ and cavitation pressure $p_\mathrm{cav} = 0\,\mathrm{Pa}$. The pressure profile is shown in Fig.~\ref{fig:twin_pslider1D_sahlin} which agrees well with the result of Ref.~\cite{sahlin2007_cavitation}.

\subsection{Flat channel with roughness}

Surface roughness plays an important role in lubrication, since small-scale height variations can lead to non-zero load bearing capacities even when the contacting surfaces are macroscopically flat, which was first observed experimentally by \citet{hamilton1966_theory}. Hence, we benchmark the presented numerical scheme on geometries that mimic surface roughness. Again, we use an example presented in Ref.~\cite{sahlin2007_cavitation}, with a height profile given by 
\begin{align}
h(x) =
\begin{cases}
h_0 +a\sin^2\left(\frac{4 \pi n}{L} \left(\frac{L}{2}-x\right)\right), & 0\leq x < L/2, \\
h_0, & L/2 \leq x \leq L.
\end{cases}
\end{align}

\begin{figure}
    \centering
    \includegraphics{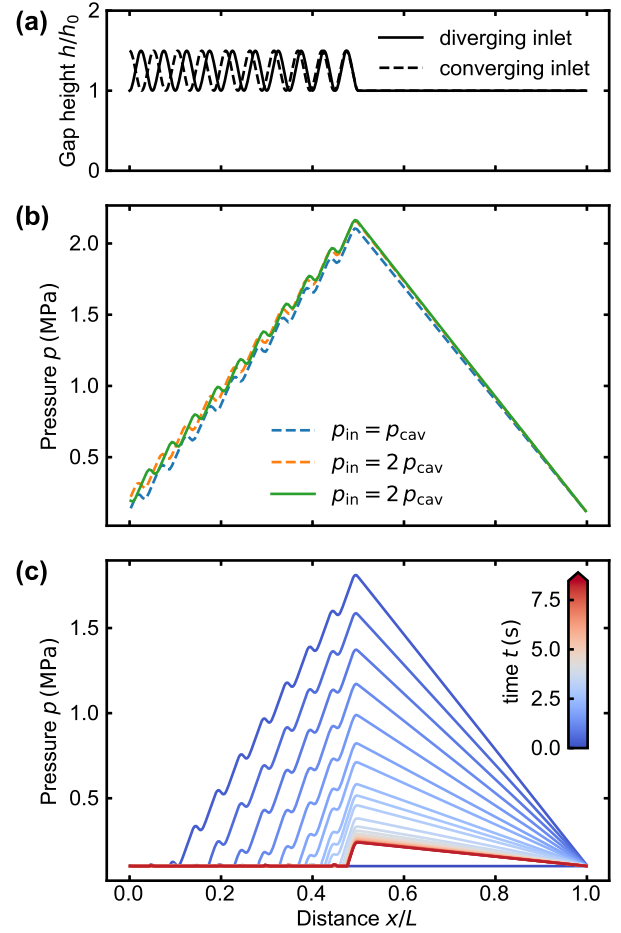}
    \caption{Geometry of the flat channel with $\sin^2$-roughness (a), corresponding pressure profiles for starved converging inlet (dashed blue line), pressurized converging inlet (dashed orange line), and pressurized diverging (solid green line) inlet (b), and graphical representation of the time evolution of the pressure profile for a starved diverging inlet (c).}
    \label{fig:rough_slider1D_sahlin}
\end{figure}

We study two realizations of this profile, with amplitude parameter $a=h_0 / 2$ and two different frequencies. The number of periods $n$ affects the gap height at the inlet, where $n=5$ leads to a diverging inlet and $n=4.75$ to a converging inlet, as shown in Fig.~\ref{fig:rough_slider1D_sahlin}a. We perform simulations for gap height $h_0 = 10\,\mathrm{\mu m}$, bearing length $L = 0.1\,\mathrm{m}$, sliding speed $U = 0.25\,\mathrm{m/s}$ and the fluid's viscosity is $\eta = 0.04\,\mathrm{Pa\,s}$. The resulting pressure profiles do not only depend on the gap height and geometry of the inlet, but also on the pressure boundary condition.

In this example, we use both a pressurized inlet ($p_\mathrm{in}=2p_\mathrm{cav}$) and a starved inlet ($p_\mathrm{in}=p_\mathrm{cav}$) with $p_\mathrm{cav}=10^5\,\mathrm{Pa}$. The former case leads to a pressure profile with load bearing capacity for both type of geometries as can be seen in Fig~\ref{fig:rough_slider1D_sahlin}b. Also, if the inlet geometry is converging and the ambient pressure is the cavitation pressure, a load bearing capacity can be achieved, since incoming fluid is immediately compressed and further surface corrugations do not lead to cavitation. The obtained pressure profiles are similar for all three cases and agree very well with the results of Ref.~\cite{sahlin2007_cavitation}.

The last case, a starved inlet with diverging geometry, is different from the previous ones as it does not generate a load bearing pressure profile. However, the results of the transient simulation reveal that initially a substantial pressure build-up is produced, which is similar to the other three situations. With increasing time, a small cavitation zone grows from the inlet into the domain and the pressure peak decreases. Furthermore, as shown in the time evolution of the pressure profile in Fig.~\ref{fig:rough_slider1D_sahlin}c, the velocity with which the cavitation zone expands decreases until it almost comes to a stand still at $t = 8\,\mathrm{s}$, with a remaining non-zero load-bearing capacity. 

\subsection{Flat channel with heterogeneous wall slip}\label{sec:slip}

Surface slip reduces friction and the possibility to control the surface chemistry \citep{xia1998_soft} opens a wide range of applications in lubrication, e.\,g. by tailoring the confining walls of a flat channel into slipping and sticking domains \citep{belyaev2010_effective, cieplak2006_nanoscale}.
Here, we study a flat channel of length $L=2\lambda$ and height $h$ with heterogeneous wetting properties at the top surface as reported in Ref.~\citep{savio2016_boundary}. The lower surface sticks and is sheared at constant velocity $U_1$. For better comparison, we here present the results for the pressure in dimensionless form, i.\,e. $\tilde{p} = p/p_0$, with $p_0 = \eta U_1 \lambda / h^2$. The sticking and slipping domains have equal length $\lambda$, and periodic boundary conditions are applied in the $x$- and $y$-direction (infinitely long bearing).

\begin{figure}
    \centering
    \includegraphics{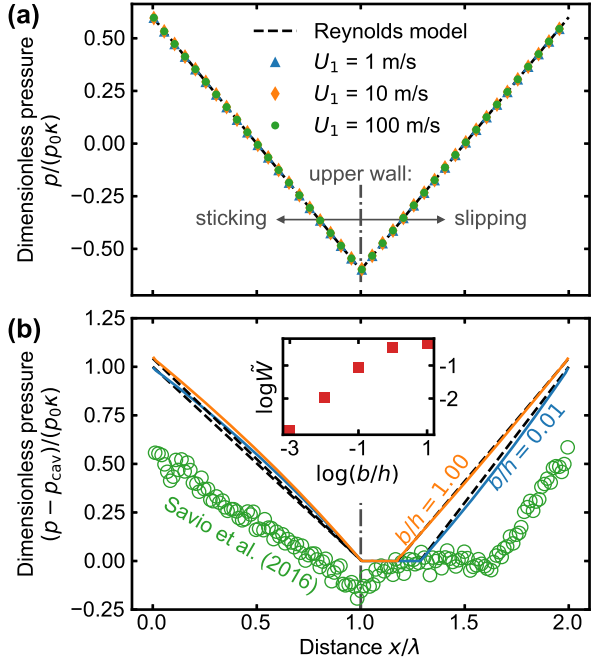}
    \caption{Non-dimensional pressure distributions for a flat channel with heterogeneous wall slip and comparison to a Reynolds model including slip. A nearly incompressible equation of state (EOS) is used in (a), which allows the pressure to go below zero. The pressure profiles for the pentane system from Ref.~\citep{savio2016_boundary} obtained from Molecular Dynamics (MD) simulations, as well as continuum results for two different slip lengths are shown in (b), where the van der Waals equation with a cutoff at $P_\mathrm{cav}=1\,\mathrm{MPa}$ is employed to model cavitation. The inset in (b) shows the non-dimensional load bearing capacity for various slip lengths.}
    \label{fig:hetslip_channel}
\end{figure}

The obtained pressure profiles for three different sliding velocities are compared to a Reynolds model including wall slip for an incompressible fluid in Fig.~\ref{fig:hetslip_channel}a. Since, our method is generally designed for compressible fluids, we arbitrarliy choose an extremely stiff EOS to mimic incompressibility. Here, we use again the Dowson-Higginson EOS and choose parameters ensuring that deviations from the reference density are negligibly small. The Reynolds model predicts a pressure gradient
\begin{align}\label{eq:gradP_slip}
    \frac{\partial p}{\partial x} = 
    \begin{cases}
    -\frac{6\eta U_1}{h^2}\left(\frac{b\lambda_2}{\lambda_1(h + 4b) + \lambda_2(h+b)}\right),& 0 \leq x \leq \lambda_1\\
    \frac{6\eta U_1}{h^2}\left(\frac{b\lambda_1}{\lambda_1(h + 4b) + \lambda_2(h+b)}\right),&L - \lambda_2 \leq x \leq L,
    \end{cases}
\end{align}
with $\lambda_1$ and $\lambda_2$ being the size of the sticking and slipping domain respectively. Thus, for $\lambda_1 = \lambda_2 = L/2$, the non-dimensional pressure gradient reads
\begin{align}\label{eq:gradP_slip_nondim}
    \frac{\partial \tilde{p}}{\partial \tilde{x}} = 
    \begin{cases}
    -6\kappa/5 , &0 \leq \tilde{x} \leq 1, \\
    6\kappa/5 , &1 \leq \tilde{x} \leq 2,
    \end{cases}
\end{align}
with $\tilde{x}=x/\lambda$ and $\kappa=5b/(2h + 5b)$. The dimensionless pressure profiles normalized by the wall-slip quantifying parameter $\kappa$ agree perfectly with the prediction of Eq.~\eqref{eq:gradP_slip_nondim}. However, both the analytical model and the numerical results predict unphysical negative pressures, and a non-zero load bearing capacity in symmetric systems with periodic boundary conditions can only be achieved by considering cavitation.

Therefore, we proceed with parameters that represent the pentane system studied with molecular dynamics (MD) simulations in Ref.~\citep{savio2016_boundary}. The channel has length $L=143.9\,\mathrm{nm}$, height $h=5.5\,\mathrm{nm}$ and sliding velocity $U_1=10\,\mathrm{m/s}$. We use the van der Waals equation as EOS
\begin{align}
    p(\rho) = \frac{RT\rho}{M -b\rho} - \frac{a\rho^2}{M^2},
\end{align}
with $a=1.926\,\mathrm{Pa\,m^6/mol^2}$,  $b=1.46\cdot10^{-4}\,\mathrm{m^3/mol}$, $M=72.15\,\mathrm{g/mol}$, $T=303\,\mathrm{K}$, and the shear viscosity is $\eta=0.3\,\mathrm{mPas}$. To avoid negative pressure and the unphysical increase of pressure with volume in the subcritical van der Waals loop, we use the aforementioned EA approach with a cutoff at $P_\mathrm{cav}=1\,\mathrm{MPa}$, which agrees with the external pressure applied in the MD simulations. Note, that in case of the van der Waals equation, the EA approach is similar to the well-known Maxwell construction \citep{clerk-maxwell1875_dynamical}, except that our choice of saturation pressure is not based on the equal area rule.

The dimensionless pressure profiles for the pentane system are shown in Fig.~\ref{fig:hetslip_channel}b for two different slip lengths. For both systems a cavitation zone forms directly at the transition from stick to slip behavior at the top wall and the size of the cavitation zone decreases with increasing slip length. We extract the size of the full film regions $\lambda_1$ and $\lambda_2$ from our simulation to compute the Reynolds solution according to Eq.~\eqref{eq:gradP_slip}. The analytical and numerical pressure profiles agree reasonably well. The slight curvature of the numerical pressure profiles can be accounted to the compressibility of the van der Waals equation in contrast to the incompressible analytical result. 

The inset of Fig.~\ref{fig:hetslip_channel}b shows the non-dimensional load bearing capacity
\begin{align}
    \tilde{W} = \frac{W}{p_0 L} = \frac{1}{p_0 L}\int_0^L p(x)\,\mathrm{d}x
\end{align}
for various slip lengths. The simulation results suggest that with increasing stick-slip contrast the system's ability to support an external load increases linearly for small slip lengths $b<<h$ until it reaches a stable value as  $b$ is in the order of the gap height or larger, which agrees with the analytical Reynolds model.

The MD simulations of \citet{savio2016_boundary} predict a substantially larger cavitation zone, and, consequently, lower pressure excursion in the remaining full-film region. The pressure falls below the external pressure in front of the stick-slip boundary, where it reaches its minimum before it goes to zero in the cavitated zone. Such complex behavior governed by the nucleation, lifetime and collapse of vapor bubbles cannot be addressed with simple macroscopic cavitation models.

This example foreshadows the potential application of the presented solver in multiscale frameworks, such as the combined use of MD simulations and continuum methods. Molecular interactions govern effects such as the nucleation of cavitation bubbles or local variations of surface slip, that are important to accurately describe lubrication phenomena on experimental length and time scales.

\subsection{2D pocket slider with non-Newtonian fluid}\label{sec:res_non-newton}

\begin{figure}
    \centering
    \includegraphics{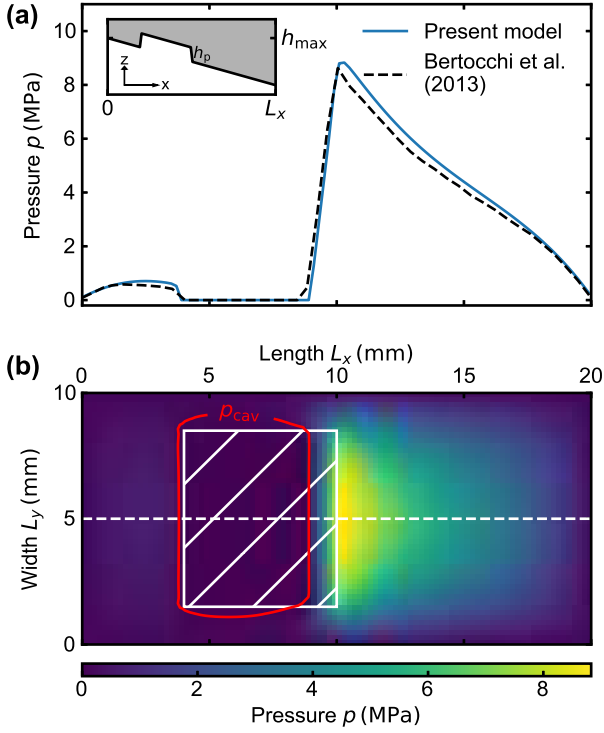}
    \caption{Center line (a) and two-dimensional (b) pressure profile for the pocket slider with cavitation and comparison to \citet{bertocchi2013_fluid}. The solid red line in (b) represents the boundary of the cavitated region ($p_\mathrm{cav}=0.1\,\mathrm{MPa}$). The dashed white line and the hatched rectangle correspond to the center line and pocket position, respectively.}
    \label{fig:pocket2D_pressure}
\end{figure}

The previous numerical examples considered only one-dimensional problems. We conclude the validation of the scheme with a two-dimensional example of a microtextured inclined slider, that has been previously used in various studies \cite{giacopini2010_massconserving, bertocchi2013_fluid, profito2015_general}. The geometry is similar to the first numerical example with $h_\mathrm{max} = 1\,\mathrm{\mu m}$, slope $s = 5 \cdot 10^{-6}$, length $L_x = 20\,\mathrm{mm}$ and width $L_y = 10\,\mathrm{mm}$. The pocket has a depth of $h_\mathrm{p} = 0.4\,\mathrm{\mu m}$, a length $l = 6\,\mathrm{mm}$, a width $w = 7\,\mathrm{mm}$ and is located at a distance $c = 4\,\mathrm{mm}$ from the inlet in $x$-direction and centered in $y$-direction. Additionally, non-Newtonian fluid behavior is described by Barus's law, Eq.~\eqref{eq:barus}, with piezoviscosity coefficient $\alpha = 12 \,\mathrm{GPa^{-1}}$  and Eyring's shear thinning law, Eq.~\eqref{eq:eyring}, with $\tau_0 = 5\,\mathrm{MPa}$. The centerline and full two-dimensional pressure profile can be seen in Fig.~\ref{fig:pocket2D_pressure}a and \ref{fig:pocket2D_pressure}b, respectively. The center line pressure distribution agrees with the result of Ref.~\cite{bertocchi2013_fluid}, which was obtained using a complementarity formulation of the cavitation problem. The relative difference of the pressure maxima is less than $3\,\%$.

\section{Conclusion}

In this paper, a height-averaged formulation of the Navier-Stokes equation for confined fluids has been derived. The time-explicit finite volume scheme to solve the governing equation involving geometrical source terms has been tested and it has been shown that various results from established solution schemes can be reproduced. The numerical tests included common lubrication problems for compressible fluids with or without consideration of inertial effects, mass-conserving cavitation, wall slip, as well as non-Newtonian fluids. The explicit solution algorithm allows to study dynamic problems, such as the time evolution of lubricant flow between geometrically or chemically structured surfaces. The power of our approach lies in the fact that any type of constitutive expression can be implemented straightforwardly. This allows testing of new constitutive models and may be useful in multiscale-simulations were fixed-form constitutive relations are not available.

\appendix

\section*{Acknowledgements}
The authors gratefully acknowledge support by the German Research Foundation (DFG) through GRK 2450, and the European Research Council (ERC) through Starting Grant 757343. Calculations were carried out on bwForCluster NEMO at the University of Freiburg (DFG grant INST 39/963-1 FUGG).

\renewcommand{\theequation}{A.\arabic{equation}}
\section*{Appendix: Viscous stress tensor components for laminar flow}\label{sec:appendix}

The components for the averaged viscous stress tensor are given by:
\begin{subequations}\label{eq:tau_avg}
\begin{align}
\bar{\tau}_{xx} &=  
-\Bigl\{\bigl(\zeta + \frac{4}{3}\eta\bigr)\bigl[(U\rho - 3 j_x)h^2+3b(U\rho-2j_x)h \nonumber \\
&\qquad + 6b^2(U\rho - j_x)\bigr]\partial_x h \nonumber \\
&\qquad +\bigl(\zeta - \frac{2}{3}\eta\bigr)\bigl[(V\rho - 3 j_y)h^2 + 3b(V\rho - 2j_y)h \nonumber \\
&\qquad+6b^2(V\rho -j_y)\bigr]\partial_y h\Bigr\} \Bigr/
h\rho(h + 4b)(h+b),\\
\bar{\tau}_{yy} &=  
-\Bigl\{\bigl(\zeta - \frac{2}{3}\eta\bigr)\bigl[(U\rho - 3 j_x)h^2+3b(U\rho-2j_x)h \nonumber \\
&\qquad + 6b^2(U\rho - j_x)\bigr]\partial_x h \nonumber \\
&\qquad +\bigl(\zeta + \frac{4}{3}\eta\bigr)\bigl[(V\rho - 3 j_y)h^2 + 3b(V\rho - 2j_y)h \nonumber \\
&\qquad+6b^2(V\rho -j_y)\bigr]\partial_y h\Bigr\} \Bigr/
h\rho(h + 4b)(h+b),\\
\bar{\tau}_{zz} &=  
-\bigl(\zeta - \frac{2}{3}\eta\bigr)\Bigl\{\bigl[(U\rho - 3 j_x)h^2+3b(U\rho-2j_x)h \nonumber \\
&\qquad + 6b^2(U\rho - j_x)\bigr]\partial_x h \nonumber \\
&\qquad +\bigl[(V\rho - 3 j_y)h^2 + 3b(V\rho - 2j_y)h \nonumber \\
&\qquad+6b^2(V\rho -j_y)\bigr]\partial_y h\Bigr\} \Bigr/
h\rho(h + 4b)(h+b),\\
\bar{\tau}_{xy} &=
-\eta \Bigl\{\bigl[(V\rho - 3 j_y)h^2 + 3b(V\rho - 2j_y)h \nonumber \\
&\qquad+6b^2(V\rho -j_y)\bigr] \partial_x h \nonumber \\
&\qquad +\bigl[(U\rho - 3 j_x)h^2+3b(U\rho-2j_x)h \nonumber \\
&\qquad + 6b^2(U\rho - j_x)\bigr]\partial_y h\Bigr\} \Bigr/
h\rho(h + 4b)(h+b),\\
\bar{\tau}_{xz} &= 
-\frac{\eta (U \rho h + 6 \rho U b - 6 j_x b)}{h\rho(h+4b)}, \\
\bar{\tau}_{yz} &= 
-\frac{\eta (V \rho h + 6 \rho V b - 6 j_y b)}{h\rho(h+4b)}.
\end{align}
\end{subequations}
At $z=h_1$, the only non-vanishing stress components are:
\begin{subequations}\label{eq:tau_bottom}
\begin{align}
    \tau_{xz}\Bigl|_{z=h_1} &= -\frac{2\eta (2U\rho h + 6 U \rho b - 3 h j_x - 6 j_x b)}{h\rho(h+4b)},\\
    \tau_{yz}\Bigl|_{z=h_1} &= -\frac{2\eta (2V\rho h + 6 V \rho b - 3 h j_y - 6 j_y b)}{h\rho(h+4b)}.
\end{align}
\end{subequations}
Viscous stress tensor components at the top wall ($z=h_2$) are given by:
\begin{subequations}\label{eq:tau_top}
\begin{align}
\tau_{xx}\Bigl|_{z=h_2} &=  
-2\Bigl\{\bigl(\zeta + \frac{4}{3}\eta\bigr)\bigl[(U\rho - 3 j_x)h^2+3b(U\rho-2j_x)h \nonumber \\
&\qquad + 6b^2(U\rho - j_x)\bigr]\partial_x h \nonumber \\
&\qquad +\bigl(\zeta - \frac{2}{3}\eta\bigr)\bigl[(V\rho - 3 j_y)h^2 + 3b(V\rho - 2j_y)h \nonumber \\
&\qquad+6b^2(V\rho -j_y)\bigr]\partial_y h\Bigr\} \Bigr/
h\rho(h + 4b)(h+b),\\
\tau_{yy}\Bigl|_{z=h_2} &=  
-2\Bigl\{\bigl(\zeta - \frac{2}{3}\eta\bigr)\bigl[(U\rho - 3 j_x)h^2+3b(U\rho-2j_x)h \nonumber \\
&\qquad + 6b^2(U\rho - j_x)\bigr]\partial_x h \nonumber \\
&\qquad +\bigl(\zeta + \frac{4}{3}\eta\bigr)\bigl[(V\rho - 3 j_y)h^2 + 3b(V\rho - 2j_y)h \nonumber \\
&\qquad+6b^2(V\rho -j_y)\bigr]\partial_y h\Bigr\} \Bigr/
h\rho(h + 4b)(h+b),\\
\tau_{zz}\Bigl|_{z=h_2} &=  
-2\bigl(\zeta - \frac{2}{3}\eta\bigr)\Bigl\{\bigl[(U\rho - 3 j_x)h^2+3b(U\rho-2j_x)h \nonumber \\
&\qquad + 6b^2(U\rho - j_x)\bigr]\partial_x h \nonumber \\
&\qquad +\bigl[(V\rho - 3 j_y)h^2 + 3b(V\rho - 2j_y)h \nonumber \\
&\qquad+6b^2(V\rho -j_y)\bigr]\partial_y h\Bigr\} \Bigr/
h\rho(h + 4b)(h+b),\\
\tau_{xy}\Bigl|_{z=h_2} &=
-2 \eta \Bigl\{\bigl[(V\rho - 3 j_y)h^2 + 3b(V\rho - 2j_y)h \nonumber \\
&\qquad+6b^2(V\rho -j_y)\bigr] \partial_x h \nonumber \\
&\qquad +\bigl[(U\rho - 3 j_x)h^2+3b(U\rho-2j_x)h \nonumber \\
&\qquad + 6b^2(U\rho - j_x)\bigr]\partial_y h\Bigr\} \Bigr/
h\rho(h + 4b)(h+b),\\
\tau_{xz}\Bigl|_{z=h_2} &= 
-\frac{2\eta (U\rho - 3 j_x)}{\rho(h+4b)}, \\
\tau_{yz}\Bigl|_{z=h_2} &= 
-\frac{\eta (V \rho - 3 j_y)}{\rho(h+4b)}.
\end{align}
\end{subequations}


\end{document}